\documentclass[preprint,12pt]{elsarticle}

\usepackage{epsfig}

\usepackage{amssymb}

\biboptions{square}

\journal{Commun. Nonlinear Sci. Numer. Simulat.}

\makeatletter

\newcommand{\refeq}[1]{(\ref{#1})}

\newcommand{\cF}{{\mathcal F}}

\newcommand{\cI}{{\mathcal I}}

\newcommand{\cL}{{\mathcal L}}

\newcommand{\cP}{{\mathcal P}}
\newcommand{\cS}{{\mathcal S}}
\newcommand{\cT}{{\mathcal T}}

\newcommand{\cZ}{{\mathcal Z}}

\newcommand{\RR}{{\mathbb R}}

\newcommand{\rmb}{{\mathrm{b}}}
\newcommand{\rmd}{{\mathrm{d}}}

\newcommand{\rmO}{{\mathrm{O}}}

\begin{document}

\begin{frontmatter}



\title{A symplectic, symmetric algorithm for spatial evolution of particles
       in a time-dependent field\tnoteref{tit1}}
\tnotetext[tit1]{\copyright{2011} The authors. Reproduction of this article,
in its entirety, for noncommercial purposes is permitted.}

\author[adr1,adr2]{A.~Ruzzon}
\ead{alberto.ruzzon@igi.cnr.it}
\author[adr1]{Y.~Elskens\corref{cor1}}
\ead{yves.elskens@univ-provence.fr}
\author[adr1]{F.~Doveil}
\ead{fabrice.doveil@univ-provence.fr}
\cortext[cor1]{corresponding author}
\address[adr1]{Equipe turbulence plasma, case 321, PIIM,
                 UMR 6633 CNRS -- Aix-Marseille universit\'e, \\
                 campus Saint-J\'er\^ome,
                 FR-13397 Marseille cedex 13}
\address[adr2]{Consorzio RFX, Corso Stati Uniti 4,
               IT-35127 Padova, Italy}

\begin{abstract}
   A symplectic, symmetric, second-order scheme is constructed
    for particle evolution in a time-dependent field with a fixed spatial step.
    The scheme is implemented in one space dimension and tested,
    showing excellent adequacy to experiment analysis. \\
\end{abstract}



\end{frontmatter}


\section{Introduction}
\label{intro}

Particle motion in a space-time dependent field is a classical fundamental
problem of
dynamics. It is generally well solved in most settings under most kinds of
requirements.
However, most solutions deal with formulations of the dynamics where the
independent
variable is time, viz.~they aim at computing a function $y$ such that the motion
reads
$x = y(t)$. Yet in some settings one actually describes motion by the reciprocal
function $\tau = y^{-1}$, so that the motion reads $t = \tau(x)$.

One such instance is the propagation of electrons in a traveling wave tube,
where it is
natural to record particles when they pass at a fixed probe location, instead of
getting
a snapshot of their locations at a given time. Similar physical contexts are met
in
other particle beam devices, such as accelerators, klystrons, free electron
lasers,
electronic tubes for wave amplification,...\citep{Kartikeyan} To some extent,
this
description is somewhat analogous to eulerian descriptions of flows in
hydrodynamics.

If one were interested in the evolution of a single particle, one could merely
compute
its motion as $x = y(t)$ and deduce its ``schedule'' function $t = \tau(x)$ and
related
quantities as functions of spatial position. In this respect, many
symplectic methods are available (see e.g.\ Refs~\cite{Hairer,Leimkuhler} for an overview),
especially
for separable hamiltonians of the form $H(p,x,t) = K(p) + V(x,t)$. However, to
describe
a beam of many particles during their spatial progression, it is reasonable to
follow
them consistently in space, to generate numerical data sampled at the same
(possibly
many) space positions. It then becomes awkward to first evolve them in time and
afterwards reconstruct their progression in spatial terms by interpolations.

\begin{figure}
  \begin{center}
     \includegraphics[width=7cm,height=4.3cm]{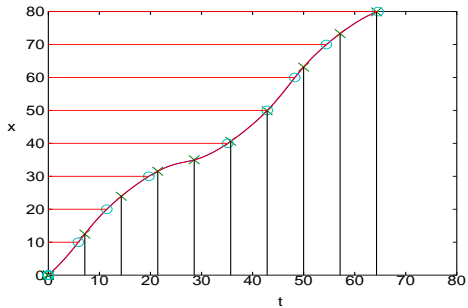}
  \end{center}
  \caption{Sketch of the algorithm principle in $(x,t)$ plot.
    The continued line is a trajectory (identical for both schemes, with small
steps).
    Crosses sample positions (every $50 \Delta t$) as obtained by leap-frog
    with constant time step,
    circles sample times (every $50 \Delta x$) as obtained by our approach
    with constant space step.
    Straight lines are guides to the eye.}
  \label{Lfig0302}
\end{figure}

For this purpose we reformulate in Section \ref{secAction} the particle
equations of
motion, using the streaming variable $x$ as independent variable (see
figure~\ref{Lfig0302}). Since the original particle dynamics is hamiltonian, we
ensure
that the new description be symplectic by first expressing the action principle
in terms
of the timetable function $\tau$. In the corresponding hamiltonian picture, the
variable
conjugate to $\tau$ is the energy $\zeta$, and the generator of motion is
momentum
$\cP$.

In Section \ref{secScheme} we stress our requirements on the scheme and
consider alternative strategies.
Then we construct a first order symplectic scheme for the particle
motion. The implicit part of the step can be performed either through algebraic
solution
of a cubic equation, or through a Newton iteration~: we compare both procedures.
Next we
construct the adjoint, first order symplectic scheme, which also requires a
Newton
iteration, and we check its accuracy. Finally, we combine the direct and adjoint
schemes
to obtain a second order symmetric, symplectic, fixed $\Delta x$ scheme.

In Section \ref{secOneWaveTest} we benchmark our algorithm by analysing the
particle
motion in the field of a single harmonic wave, viz.~we solve the pendulum motion
in a
galilean frame. Numerical simulations for realistic beam data generate beam
deformations
shown in Section \ref{secOneWaveBeam}.

Section \ref{secTwoWaves} focuses on the evolution of a beam launched in
presence of two
harmonic waves. Simulations are confronted with experimental observations of the
beam
collected at the device outlet. Special attention is paid to the reproduction of
a devil
staircase structure, characteristic of the chaotic behaviour of the system,
taking into
account the finiteness of the experimental device.

In summary, experimental data often relate to limited interaction times, while
numerical
evidences and theoretical discussions of chaotic dynamics often deal with
trajectories
followed for long times in a compact domain of phase space. The agreement of our
simulations with experimental evidence assesses the relevance of our algorithm
to such
experimental settings.

\section{Evolution with respect to space}
\label{secAction}

Rewriting the equations of motion in hamiltonian form with respect to space is
straightforward in the symplectic formalism (see Section \ref{Hamilton}).
However one may wish first a more pedestrian derivation, from the classical
action principle.

\subsection{Lagrangian viewpoint}
\label{Lagrange}

The action for a non-relativistic particle with mass $m$
moving along a one-dimensional
axis $\mathrm{O} x$ in a time dependent potential $V(x,t)$ reads
\begin{equation}
  S [y ; t_0, t_1, x_0, x_1]
  =
  \int_{t_0}^{t_1} L(y(t), \dot y(t), t) \rmd t
  \label{actionSt}
\end{equation}
where $y$ is a continuously differentiable function of time $t \in [t_0, t_1]
\subset
\RR$, subject to the constraints $y(t_0) = x_0$ and $y(t_1) = x_1$, and the dot
denotes
derivative with respect to $t$. The lagrangian is
\begin{equation}
  L(x,v,t) = \frac{m v^2}{2} - V(x,t)
  \label{defLt}  \, .
\end{equation}
In the following we restrict all trajectories to the class of strictly monotone,
increasing functions, viz.\ $\inf_{t_0 < t < t_1} \dot y(t) > 0$. For these
functions
the reciprocal function
  $\tau : [x_0, x_1] \to [t_0, t_1] : x \mapsto \tau(x)$ exists ; $\tau$ is
unique
and also strictly monotone, increasing, continuously differentiable with
\begin{equation}
  \tau'(x)
  = \frac {\rmd \tau}{\rmd x} (x)
  = \left(\frac{\rmd y}{\rmd t} (\tau(x)) \right)^{-1}
  = \frac{1}{\dot y(\tau(x))}
  \label{tauprime}
\end{equation}
where the prime denotes derivative with respect to $x$.

To rewrite \refeq{actionSt} as a space-integral, we introduce the new lagrangian
\begin{equation}
  \cL (t, u, x) := L(x, \frac{1}{u}, t) u
  \label{defLamz}
\end{equation}
so that
\begin{equation}
  \cS [\tau ; x_0, x_1, t_0, t_1]
  :=
  S [\tau^{-1} ; t_0, t_1, x_0, x_1]
  =
  \int_{x_0}^{x_1} \cL (\tau(x), \tau'(x), x) \rmd x
  \label{actionSz}  \, .
\end{equation}
It is convenient to introduce the opposite to the (usual definition of)
canonical
momentum conjugate to $\tau$,
\begin{equation}
  \zeta = - \, \frac{\partial \cL}{\partial u}
  \label{defzeta}
\end{equation}
and to perform the Legendre transform of $-\cL$, defining
\begin{equation}
  \cP (\zeta, t, x)
  := \zeta \tau' + \cL(t, \tau', x)
  \label{defP}
\end{equation}
so that in the new variables the canonical Hamilton equations read
\begin{eqnarray}
  \frac{\rmd \tau}{\rmd x}
  & = &
  \frac{\partial \cP}{\partial \zeta}
  \label{dPdt}  \, ,
  \\
  \frac{\rmd \zeta}{\rmd x}
  & = &
  - \, \frac{\partial \cP}{\partial \tau}
  \label{dPdz}  \, .
\end{eqnarray}

For the classical lagrangian \refeq{defLt} the new variables are the usual energy and
linear momentum,
\begin{eqnarray}
  \zeta
  & = &
  \frac{m}{2 {\tau'}^2} + V(x,\tau(x))
  \label{zeta2} \, ,
  \\
  \cP
  & = &
  \sqrt{2 m ( \zeta - V(x,\tau(x)) )}
  \label{cP2}  \, .
\end{eqnarray}

\subsection{Hamiltonian viewpoint}
\label{Hamilton}

The hamiltonian formulation of dynamics provides a direct path to the latter
equations.
Indeed it suffices to consider the symplectic 2-form
\begin{equation}
  \rmd \omega
  := \rmd p \rmd x - \rmd H \rmd t
\label{2form}
\end{equation}
where $p = m \dot y$ is conjugate to $x$ and $H = \zeta$ is conjugate to $t = \tau$, and
to consider $x$ as the independent variable along trajectories instead of $t$. The minus
sign we introduced in \refeq{defzeta}-\refeq{defP} ensures the usual signs in $\rmd
\omega$.

Of course, if the potential does not depend explicitly on time, $\zeta$ is a
first integral.

The above requirement, that $\dot y$ nowhere vanishes, will be strengthened below by
imposing $\cP \geq p_{\min} > 0$ for some $p_{\min}$ to ensure appropriate numerical
accuracy.

\section{Discrete time canonical transformations}
\label{secScheme}

To preserve the symplectic structure while integrating \refeq{zeta2}-\refeq{cP2}
we use
a sequence of canonical transformations. Note that \refeq{cP2} does not allow
using a
splitting method such as leap-frog \citep{Hairer,Leimkuhler}, as it is not the
sum of integrable generators.

In order to sample the spatial evolution regularly (as desired
 e.g.\ to follow many particles in parallel), we use a fixed spatial step
$\Delta x$, a strategy recommended e.g.\ by H\'enon \cite{Henon} to obtain simple and
accurate Poincar\'e maps (for a space periodic potential, the step $\Delta x$ is best
taken as a fraction of the wavelength). A fixed spatial step also avoids generating
noisy-like ``spurious frequencies'' in the simulated dynamics. Note that we could also
use a time-integrator (leap-frog or other) with an adapting time step, $\Delta t = m
\Delta x/\cP$ \cite{Henon}, but it is not manifestly symplectic as $\cP$ is not
constant. Another approach, using a fixed time step and interpolating the trajectories
for the spatial mesh, would raise the issue of constructing a symplectic interpolation
scheme (of desired order). In this work we settle for the manifestly symplectic, fixed
step approach, which can process many particles in parallel.

Our first integrator
$\cF_{\Delta x} : (\tau, \zeta, x) \mapsto (\tilde \tau, \tilde \zeta, x +
\Delta x)$
is chosen to provide an explicit, first-order approximation for the time
increment. It is generated by
\begin{equation}
  F(\tau, \tilde \zeta)
  =
  \tau \tilde \zeta + \cP(\tau, \tilde \zeta, x) \Delta x
  \label{Fint1}
\end{equation}
so that the system
\begin{eqnarray}
  \zeta
  & = &
  \tilde \zeta
  + \partial_\tau \cP (\tau, \tilde \zeta, x) \Delta x
  \label{map1zeta} \, ,
  \\
  \tilde \tau
  & = &
  \tau
  + \partial_\zeta \cP (\tau, \tilde \zeta, x) \Delta x
  \label{map1tau} \, ,
\end{eqnarray}
is a first-order, symplectic approximation to \refeq{dPdt}-\refeq{dPdz}. As the second
equation is explicit with respect to time $\tilde \tau$, the first equation is implicit
with respect to the energy $\tilde \zeta$. For the special case of momentum \refeq{cP2},
it actually leads to a cubic equation,
\begin{equation}
  \tilde \zeta^3
  - (V + 2 \zeta) \tilde \zeta^2
  + (2 V \zeta + \zeta^2) \tilde \zeta
  - V \zeta^2 - \frac{m}{2} (\partial_\tau V)^2 \Delta x^2
  = 0
  \label{can1cube} \, ,
\end{equation} which can be solved algebraically for $\tilde \zeta$. Here $V$
and $\partial_\tau V$ are computed at $(x, \tau)$. Equation~\refeq{can1cube} usually has
three real solutions, two of which being $\Delta x$-close to $\zeta$ : the relevant root
is such that $(\zeta - \tilde \zeta)\,  \partial_\tau V > 0$.

It is advantageous to express \refeq{map1zeta} in the form
 $\tilde \zeta = \zeta + (\zeta - V) \sigma$, so that
\begin{equation}
  \sigma = - a \, (1 + \sigma)^{-1/2}
  \label{eqsigma}
\end{equation}
with the single parameter
\begin{equation}
  a = \sqrt{m/2}\, (\zeta - V)^{-3/2} \, \partial_\tau V \Delta x \,
  \label{perta}
\end{equation}
calculated at $(x , \tau)$. The fixed point equation \refeq{eqsigma} is easily solved by
the Newton method, which selects the ``good'' cubic root automatically (as $a \sigma <
0$) and converges very fast, especially if $a$ is small. Analytically, this method
stresses the small parameter $a$ controlling the accuracy of our scheme : it involves a
balance of the potential evolution $\partial_\tau V$ and the spatial step $\Delta x$
against $(\zeta - V)$, viz.\ against the particle velocity $\cP/m$. For small velocity,
the algorithm deteriorates -- at worst it will miss turning points where $\cP$ changes
sign (which is forbidden by our assumptions on trajectories in the action principle).

Let $\cZ_{\Delta x}^{\mathrm C} : (\tau, \zeta, x) \mapsto (\tau, \tilde \zeta,
x)$
 and $\cZ_{\Delta x}^{\mathrm N} : (\tau, \zeta,x) \mapsto (\tau, \tilde \zeta,
x)$
denote respectively the cubic and Newton solvers. For perfectly accurate
computations, they coincide and may be denoted identically $\cZ_{\Delta x}$.
Note that
$\cZ_{\Delta x}$ is not symplectic, as
\begin{equation}
  \det D \cZ_{\Delta x} (\tau, \zeta, x)
  =
  [1 + \partial_\zeta \partial_\tau \cP (\tau, \tilde \zeta, x) \Delta x]^{-1}
  \label{detDZ} \, .
\end{equation}

With the new energy $\tilde \zeta$, \refeq{map1tau} immediately provides the new
time,
defining the map  $\cT_{\Delta x} : (\tau, \tilde \zeta,x) \mapsto (\tilde \tau,
\tilde
\zeta, x)$. This map is not symplectic either, as
\begin{equation}
  \det D \cT_{\Delta x} (\tau, \tilde \zeta, x)
  =
  1 + \partial_\zeta \partial_\tau \cP (\tau, \tilde \zeta, x) \Delta x
  \label{detDT} \, .
\end{equation}
Finally, we advance position, with $\cI_{\Delta x} : (\tau, \zeta, x) \mapsto (\tau,
\zeta, x + \Delta x)$. The resulting integration scheme $\cF_{\Delta x} = \cI_{\Delta x}
\circ \cT_{\Delta x} \circ \cZ_{\Delta x}$ is symplectic by construction, within machine
accuracy, as the planar map $\cT_{\Delta x} \circ \cZ_{\Delta x}$ is area-preserving :
$\det D (\cT_{\Delta x} \circ \cZ_{\Delta x}) = (\det D \cT_{\Delta x}) (\det D
\cZ_{\Delta x}) = 1$. It is first order only.

The variables advanced with \refeq{map1zeta} and \refeq{map1tau} are in the form
\begin{eqnarray}
\tilde \tau
 &~\cong~&
 \tau + \tau ' \Delta x + \frac{\tau ''}{2} \Delta x^2 + \rmO(\Delta x^3) \, ,
\\
 \tilde \zeta
 &~\cong~&
 \zeta + \zeta ' \Delta x + \frac{\zeta ''}{2} \Delta x^2 + \rmO(\Delta x^3) \,
,
\end{eqnarray}
where $\tau '$ and $\zeta '$ have been approximated with $- \partial_\tau \cP
(\tau, \tilde \zeta, x)$ and $\partial_\zeta \cP (\tau, \tilde \zeta, x)$. It
follows that the most influential theoretical error, in every step of
integration, is given by $\tau '' \Delta x / 2$ for $\tilde \tau$ and $\zeta ''
\Delta x / 2$ for $\tilde \zeta$. Table~\ref{tableerreur} compares the upper
estimated theoretical errors, related to $\tau ''$ and $\zeta ''$, and the
maximum real simulation errors for five particle initial velocities. Slower
particles are found affected by larger errors as expected.

\begin{table}
 \begin{center}
 \begin{tabular}{|l|l|l|l|l|}
 \hline
 $v_\mathrm{in}$ & $\delta\zeta_\mathrm{estimate}$ &
$\delta\zeta_\mathrm{simulation}$ & $\delta\tau_\mathrm{estimate}$ &
$\delta\tau_\mathrm{simulation}$ \\
 \hline
 2.5 & 0.16 & $6\cdot 10^{-4}$ & 0.02 & 0.005 \\
 \hline
 1.5 & 0.11 & $9\cdot 10^{-4}$ & 0.10 & 0.03 \\
 \hline
 0.9 & 0.10 & 0.008 & 8 & 0.5 \\
 \hline
 0.6 & 20 & 0.1 & 12 & 5 \\
 \hline
 0.4 & 3 & 0.3 & 70 & 16 \\
 \hline
 \end{tabular}
 \end{center}
 \caption{The largest single step error for five particles with different
initial velocity launched in the field of a single wave with $A =0.1$, $k =
0.2$, $v_\phi = 1$, $\phi = \pi/4$.}
 \label{tableerreur}
\end{table}

To obtain a second order, symmetric scheme, we consider the adjoint map
\citep{Hairer},
which is also symplectic, generated by the function
\begin{equation}
  F^*(\tilde \tau, \zeta)
  = \tilde \tau  \zeta - \cP(\tilde \tau, \zeta, x + \Delta x) \Delta x
  \label{Fint1a}
\end{equation}
so that
\begin{eqnarray}
  \tau
  & = &
  \tilde \tau - \partial_\zeta \cP (\tilde \tau, \zeta, x + \Delta x) \Delta x
  \label{map1atau} \, ,
  \\
  \tilde \zeta
  & = &
  \zeta - \partial_\tau \cP (\tilde \tau, \zeta, x + \Delta x) \Delta x
  \label{map1azeta} \, .
\end{eqnarray}
For momentum \refeq{cP2}, both equations involve the potential
 $V(x + \Delta x, \tilde \tau)$, which implies that one first solves
\refeq{map1atau} with respect to the new time $\tilde \tau$ by a Newton algorithm, and
then computes the new energy $\tilde \zeta$ by \refeq{map1azeta}. This defines the
symplectic map $\cF_{\Delta x}^{*} = \cZ_{\Delta x}^* \circ \cT_{\Delta x}^* \circ
\cI_{\Delta x} : (\tau, \zeta, x) \mapsto (\tilde \tau, \tilde \zeta, x + \Delta x)$.

One easily checks that $\cI_{\Delta x}$ is self-adjoint, while $\cZ_{\Delta x}^*
\circ
\cZ_{- \Delta x}$ and $\cT_{\Delta x}^* \circ \cT_{- \Delta x}$ reduce to
identity
up to machine numerical tolerance (typically $10^{-15}$).

Finally the composition
\begin{equation}
  \cF_{\Delta x}^{(2)}
  =
  \cF_{\Delta x/2}^{*}
  \circ
  \cF_{\Delta x/2}
  =
  \cZ_{\Delta x/2}^* \circ \cT_{\Delta x/2}^*
  \circ \cI_{\Delta x} \circ
  \cT_{\Delta x/2} \circ \cZ_{\Delta x/2}
  \label{map2} \,
\end{equation} is its own adjoint. It is thus symmetric and therefore second
order \citep{Hairer}.

\section{Validation : particle dynamics in a single wave}
\label{secOneWaveTest}

\begin{figure}
  \begin{center}
     \includegraphics[width=5cm,height=4cm]{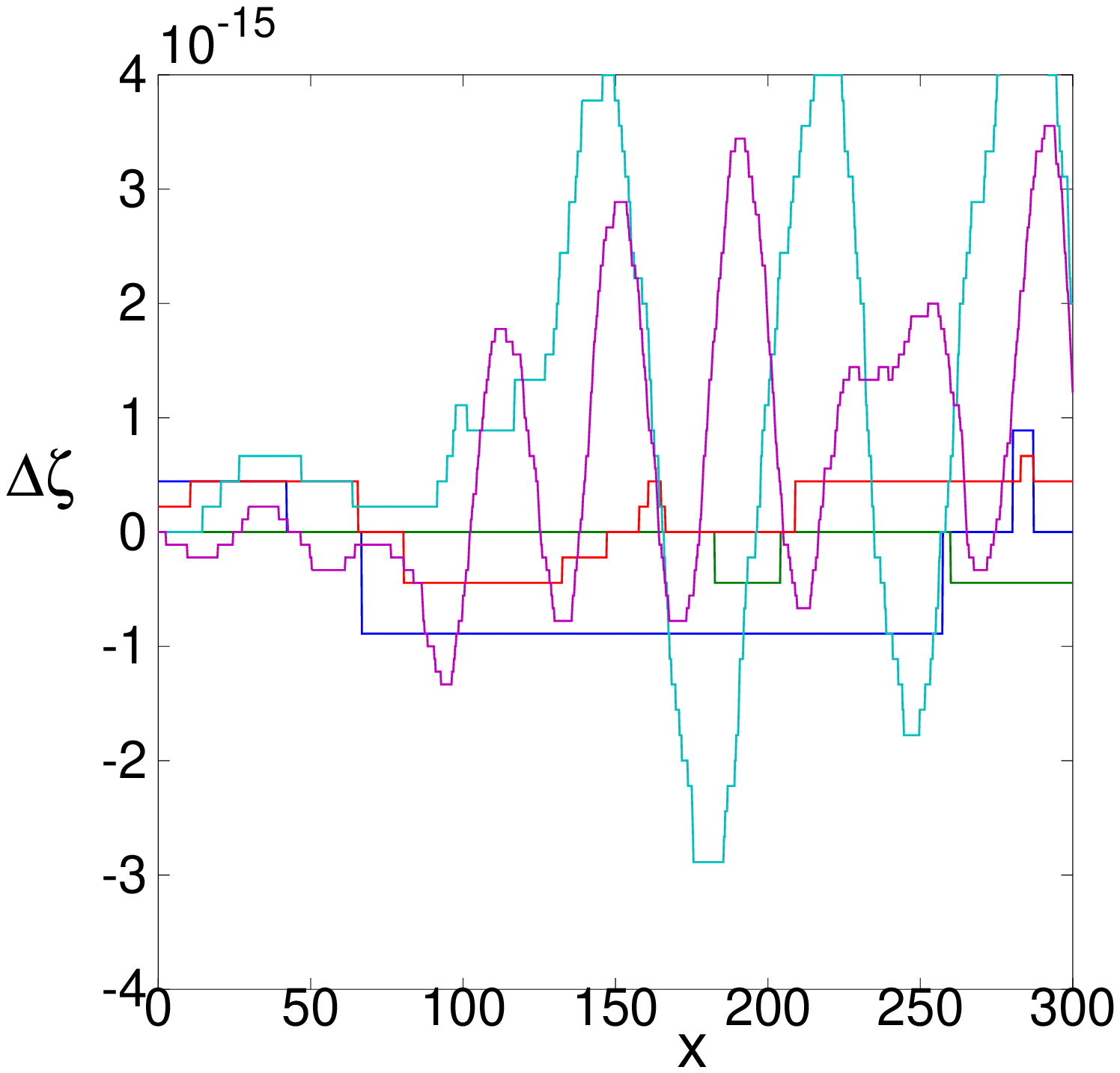}
     \includegraphics[width=5cm,height=4cm]{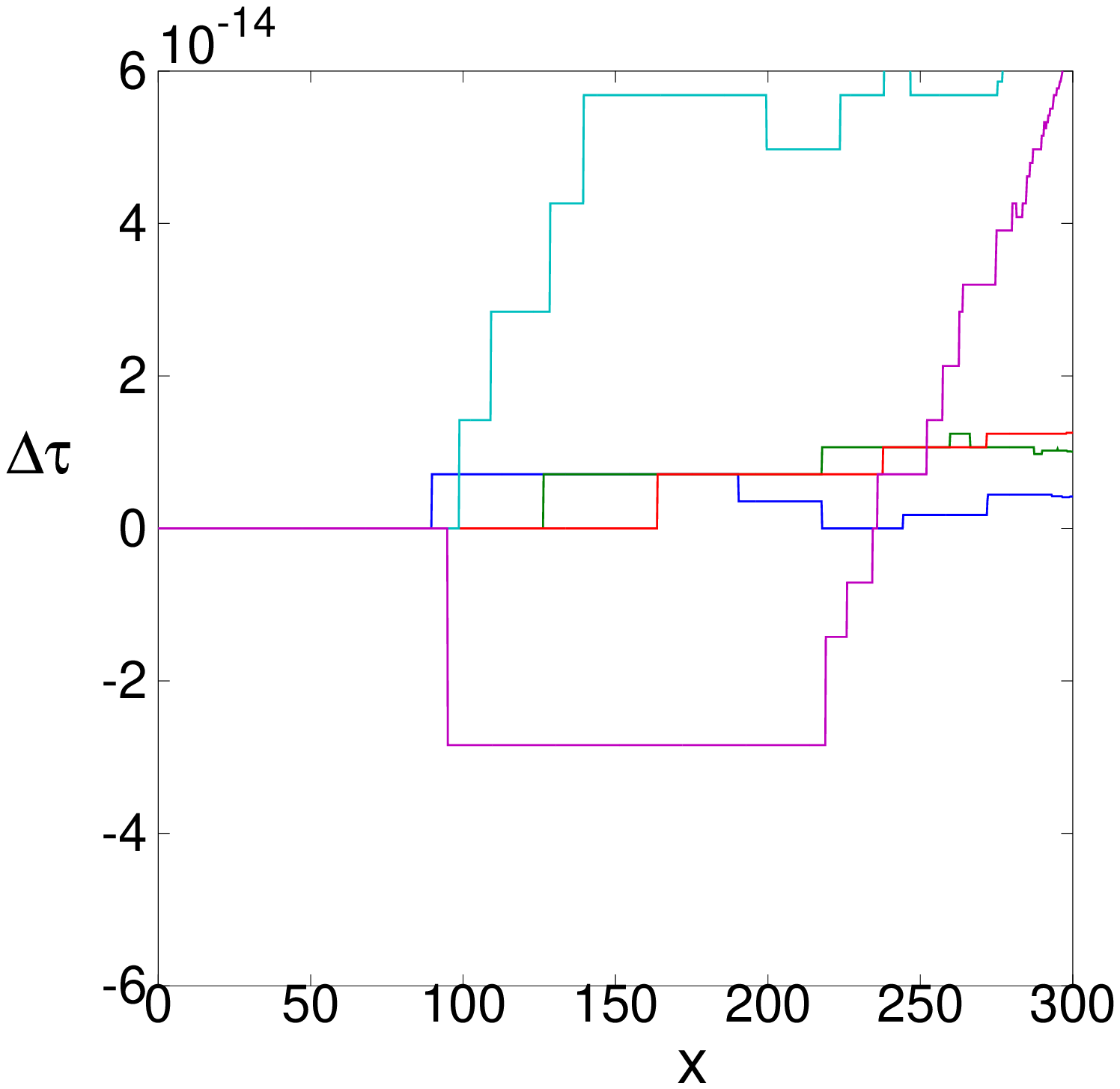}
     \includegraphics[width=5cm,height=4cm]{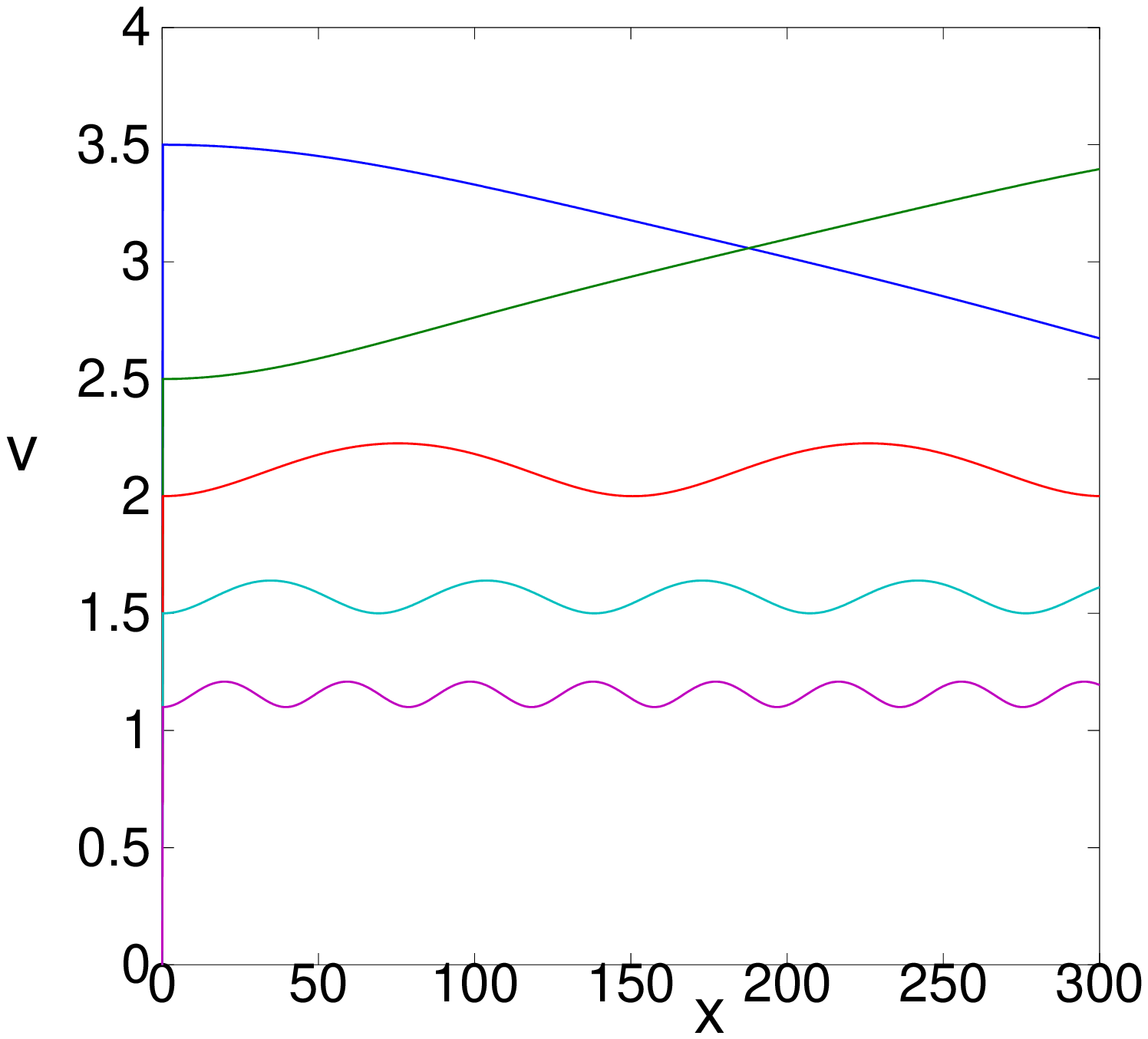}
  \end{center}
  \caption{Discrepancy, between direct evolution $\cF_{\Delta x}$ and backward
evolution
  $\cF^*_{- \Delta x}$, for energy (left panel) and arrival time (centre) as
functions of position,
  for five particles in the field of a single wave
  with $A =0.1$, $k = 0.2$, $v_\phi = 1$, $\phi = \pi$.
  Particles injected at the origin at $t=0$, with velocities $1.1$, $1.5$, $2$,
$2.5$ and $3.5$,
  move as shown on right panel.}
  \label{Lfig1roundtrip}  
\end{figure}

We test our schemes with the time dependent potential of a wave
\begin{equation}
  V(x,t) = A \cos (k x - k v t + \phi)
  \label{V1wave}
\end{equation}
where $A, k, v, \phi$ are respectively the amplitude, wavevector, phase velocity
and
phase of the wave. Rescaling energy (and amplitude), space and time enables one
to set
$m$, $k$ and $v$ to unity, and the choice of the origin of time or space
eliminates
$\phi$. Its integrability makes this dynamics a good benchmark.

\begin{figure}
  \begin{center}
     \includegraphics[width=7.5cm,height=5cm]{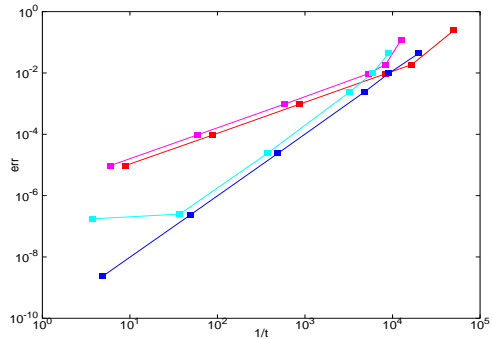}
  \end{center}
  \caption{Numerical accuracy for the first order algorithm (red, upper lines)
    and second order (blue, lower lines)
    versus calculation time per spatial unit length. Darker colour for the
Newton method solution to
    the cubic equation, lighter colour for polynomial solver
    (slightly slower than Newton).}
  \label{LfigErrScale} 
\end{figure}

The accuracy of the determination of the adjoint map is checked by iterating first
$\cF_{\Delta x}$ for $\Delta x = 0.01$ from $x = 0$ to $x = L = 300$, and then
$\cF^*_{-\Delta x}$ from $x = L$ to $x = 0$, for five particles.
Figures~\ref{Lfig1roundtrip} display the discrepancies $\Delta \zeta$ and $\Delta \tau$
as functions of $x$ for each particle and confirm that $\cF^*_{- \Delta x} = \cF_{\Delta
x}^{-1}$ to numerical accuracy. The order of the algorithms and their accuracy is
further analysed in figure~\ref{LfigErrScale}, comparing the first order and second
order schemes for the motion of a particle with initial velocity $v_\mathrm{in} = 1.5$
in the field of a wave with $A = 0.1, \phi = \pi, k = 0.2, v_\phi = 1$ over a length $L
= 100$.

\section{Beam dynamics in a single wave}
\label{secOneWaveBeam}

\begin{figure}
  \begin{center}
    \includegraphics[width=7cm,height=4.3cm]{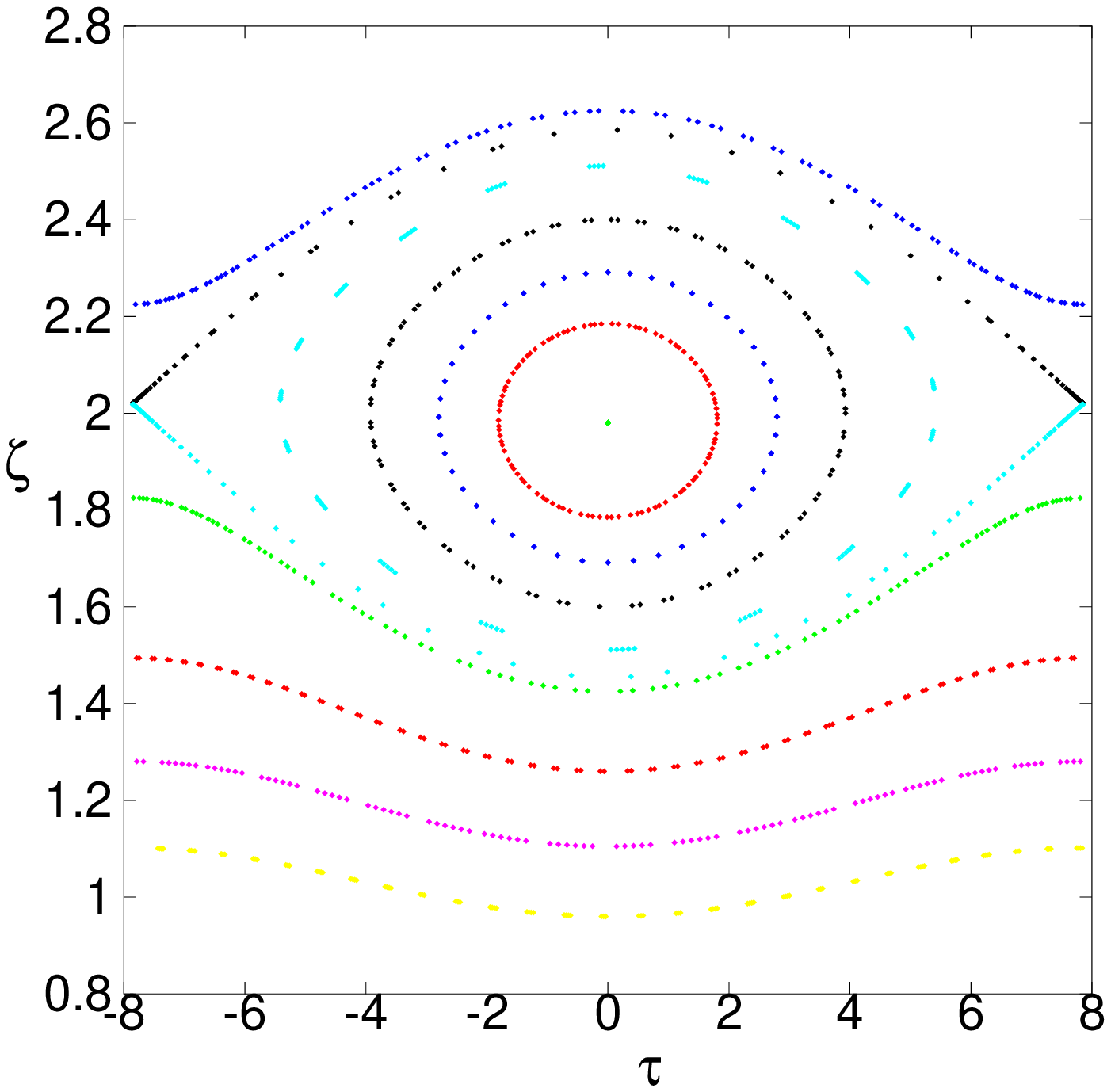}
    \includegraphics[width=7cm,height=4.3cm]{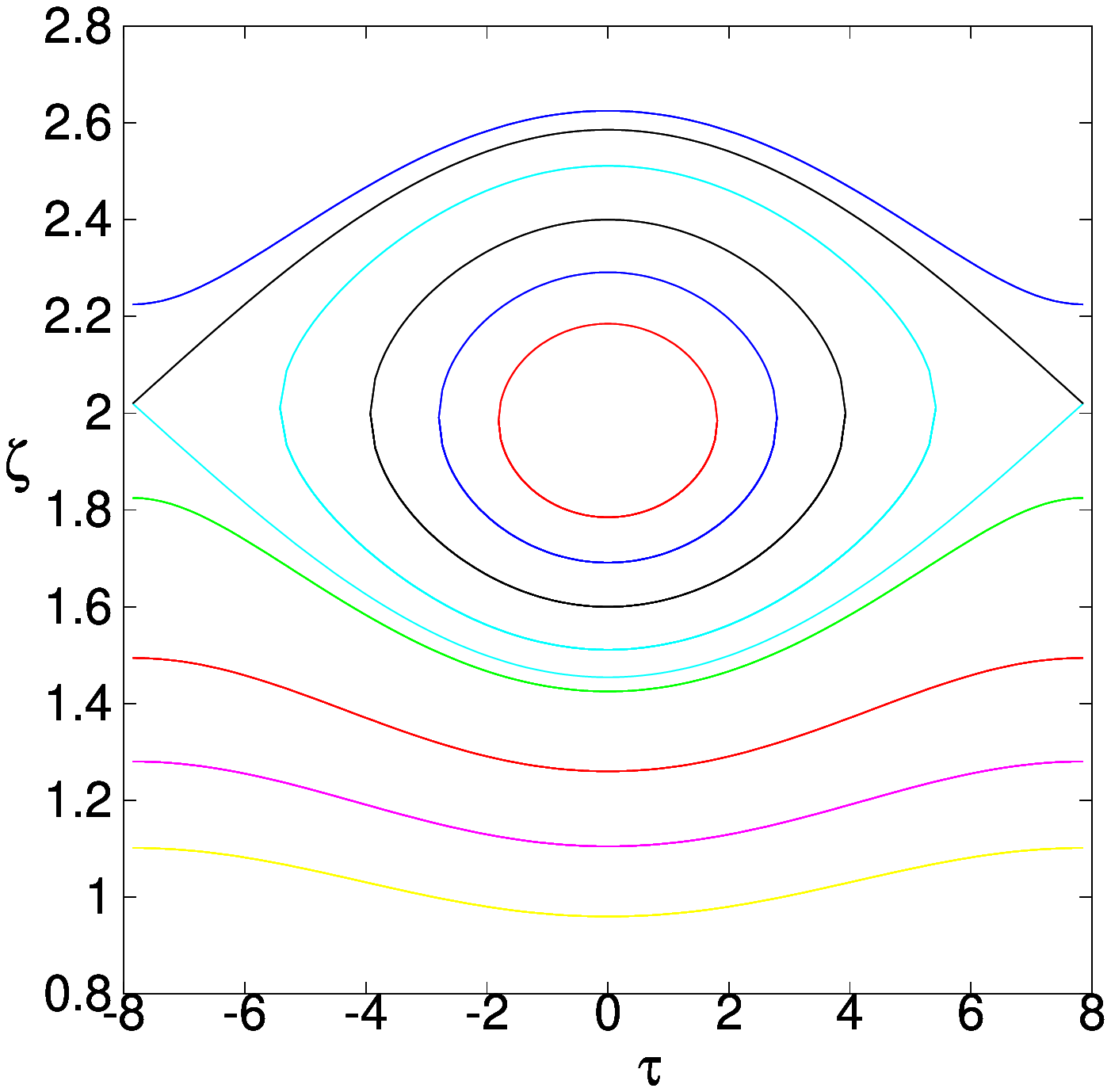}
  \end{center}
  \caption{(left) Poincar\'e section in $(\tau, \zeta)$ variables, of the return
    map after one wavelength $L$.
    Trajectories for five trapped particles injected with
    $v_\mathrm{in} = 1.8, 1.85, 2, 2.1, 2.25$,
    for five untrapped particles with
    $v_\mathrm{in} = 1.4, 1.5, 1.6, 1.7, 2.3$,
    and for one particle injected on each separatrix branch.
    Wave parameters $A = 0.02, v_\phi = 2, k = 0.2, \phi = \pi$ ;
    particle mass $m=1$.
    (right) Exact section lines \refeq{eqtraj1w}.}
\label{LfigPoinca1} 
\end{figure}

A second accuracy check is provided by the Poincar\'e section of the beam by the
positions $x \, \mathrm{mod}\, L = 0$. The return map for $(\tau, \zeta)$
variables is
symplectic. As the particle motion is integrable (it reduces to the pendulum by
a
Galileo transformation to the reference frame comoving with the wave), each
orbit must
generate section points on lines satisfying the algebraic relation
\begin{equation}
  \zeta
  = \frac{m v_\phi^2 + \bar H}{2}
  \pm v_\phi \sqrt{ 2 m \bar H - 2 m A \cos [k (x - v_\phi \tau) + \phi]}
\label{eqtraj1w}
\end{equation}
where $\bar H$ is a constant. In particular, the motion on the wave separatrix
corresponds to $\bar H = A$. For $\bar H > A$, this relation defines two
branches
for all times $\tau$, which correspond to faster or slower circulating
particles, while
for $\bar H < A$ the relation defines the upper and lower part of trapped motion
inside the wave's cat eye. Figure \ref{LfigPoinca1} shows that numerical
trajectories
perfectly reproduce these lines.

To assess the relevance of the algorithm to experiment we also follow the
deformation of a beam of electrons injected in a
single wave, e.g.\ in a traveling wave tube. As particles are accelerated or
decelerated by the wave, the beam
velocity profile is deformed. We thus inject $N$ particles at $x=0$, equally
distributed
over one time period of the wave, and plot the histogram of particle velocities
as a
function of abscissa $x$.

\begin{figure}
  \begin{center}
    \includegraphics[width=7cm,height=4.3cm]{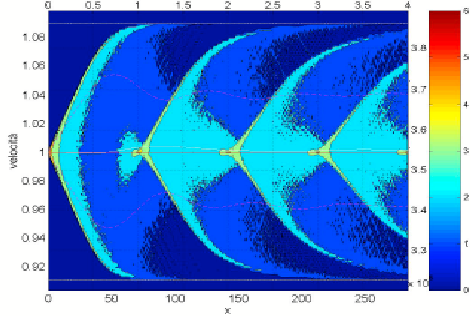}
  \end{center}
  \caption{Velocity distribution function of particles along the $x$ axis,
    injected at the wave phase velocity, $v_\rmb = v_\phi = 1$, with $m = k = 1$
    and wave amplitude $A = 0.002$.}
  \label{Lfig0403}
\end{figure}

In figure~\ref{Lfig0403} a cold beam is made of particles injected resonantly
with the
wave velocity, $v_\rmb = v_\phi = 1$, with $m = k = 1$. The wave amplitude, $A =
0.002$,
determines the bounce frequency $\omega_\rmb = \sqrt{kA}$ so that particle
oscillations
in the wave trough have a spatial period $L_\rmb = 2 \pi v_\phi / \omega_\rmb =
140.5$.
The particles bounce indeed and, in agreement with the rotating bar
approximation
\citep{MK78}, most of them reconvene every $L_\rmb / 2$. Only the ones injected
at times
close to $(\phi + 2 \pi n) / (k v_\phi) $ (for integer $n$) enter the wave close
to the
X point and follow closely the inner side of the cat eye separatrix. The time
these
particles need to overcome half a wavelength can be arbitrarily long, so that
they mark
the boundary of the wave resonant domain at velocities $v_\phi \pm 2
\sqrt{A/m}$.

\begin{figure}
 \begin{center}
  \includegraphics[width=7cm,height=4.3cm]{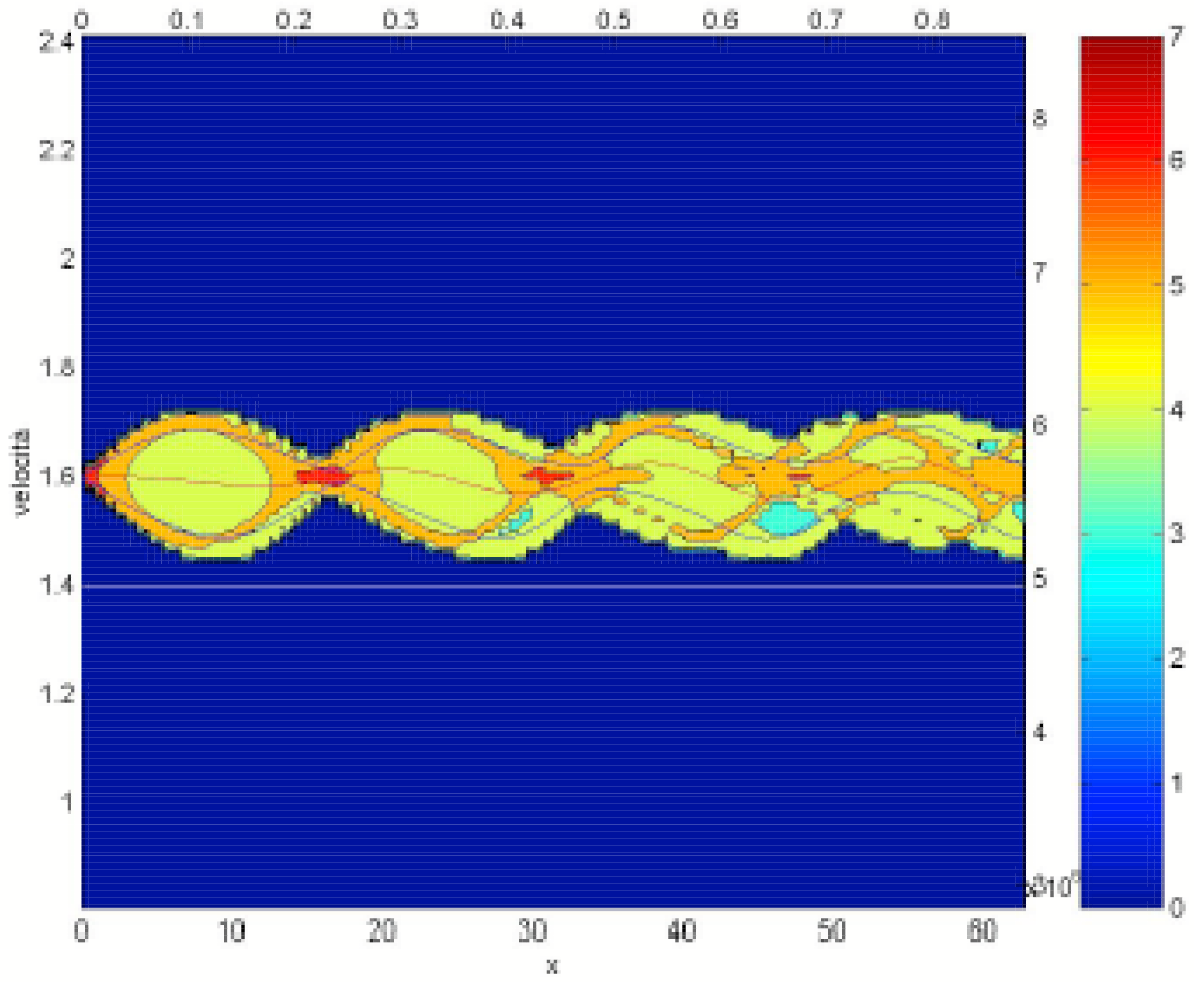}
  \includegraphics[width=7cm,height=4.3cm]{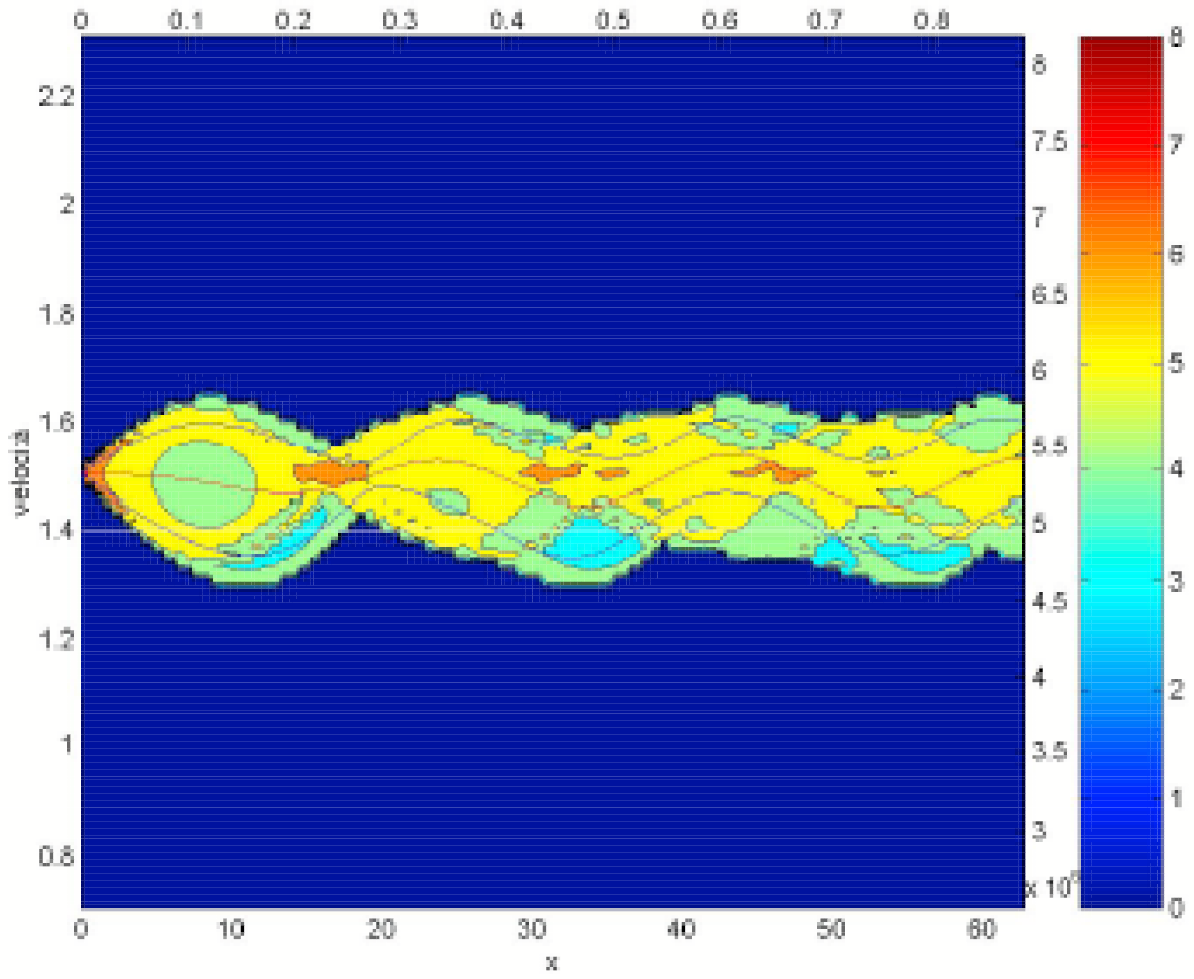}
  \includegraphics[width=7cm,height=4.3cm]{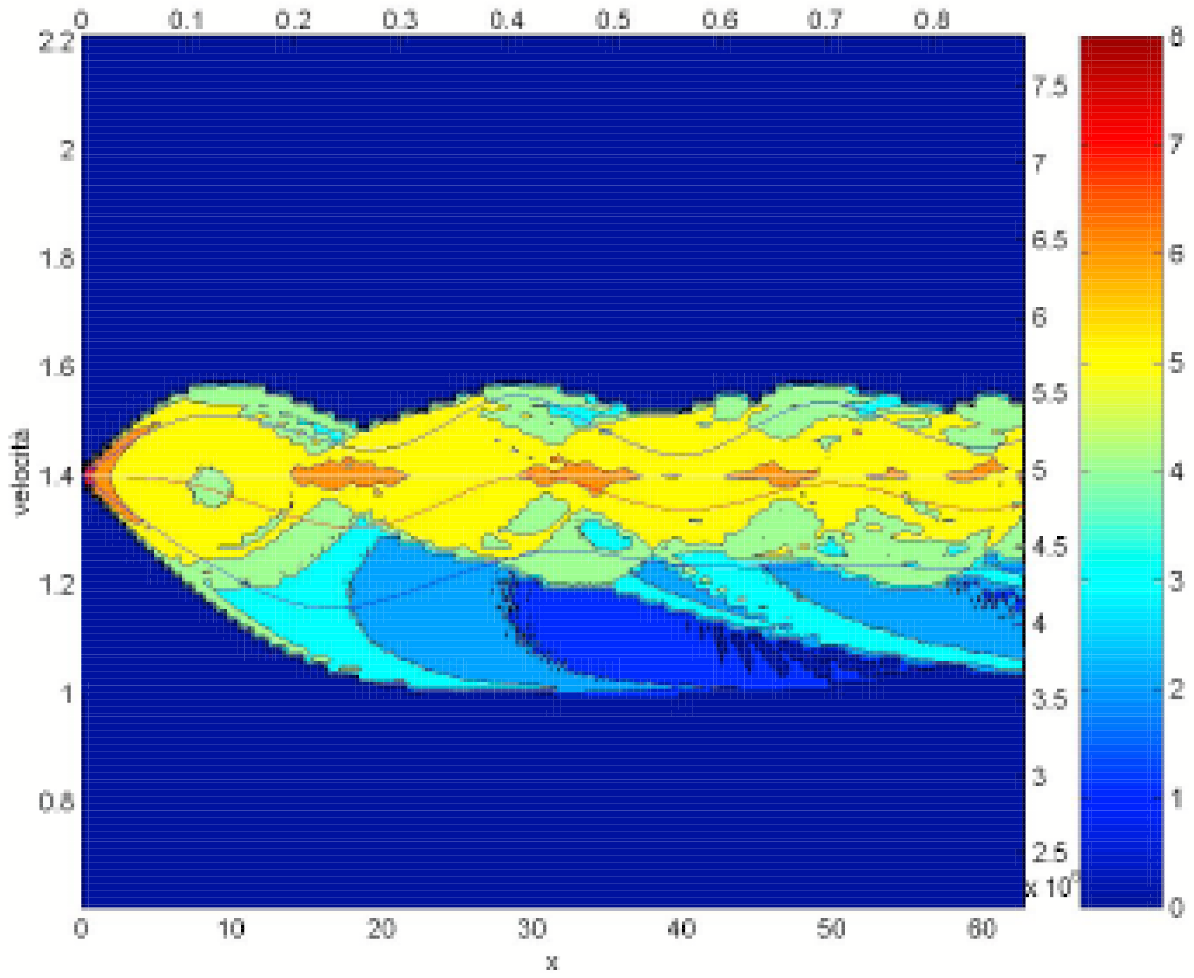}
  \includegraphics[width=7cm,height=4.3cm]{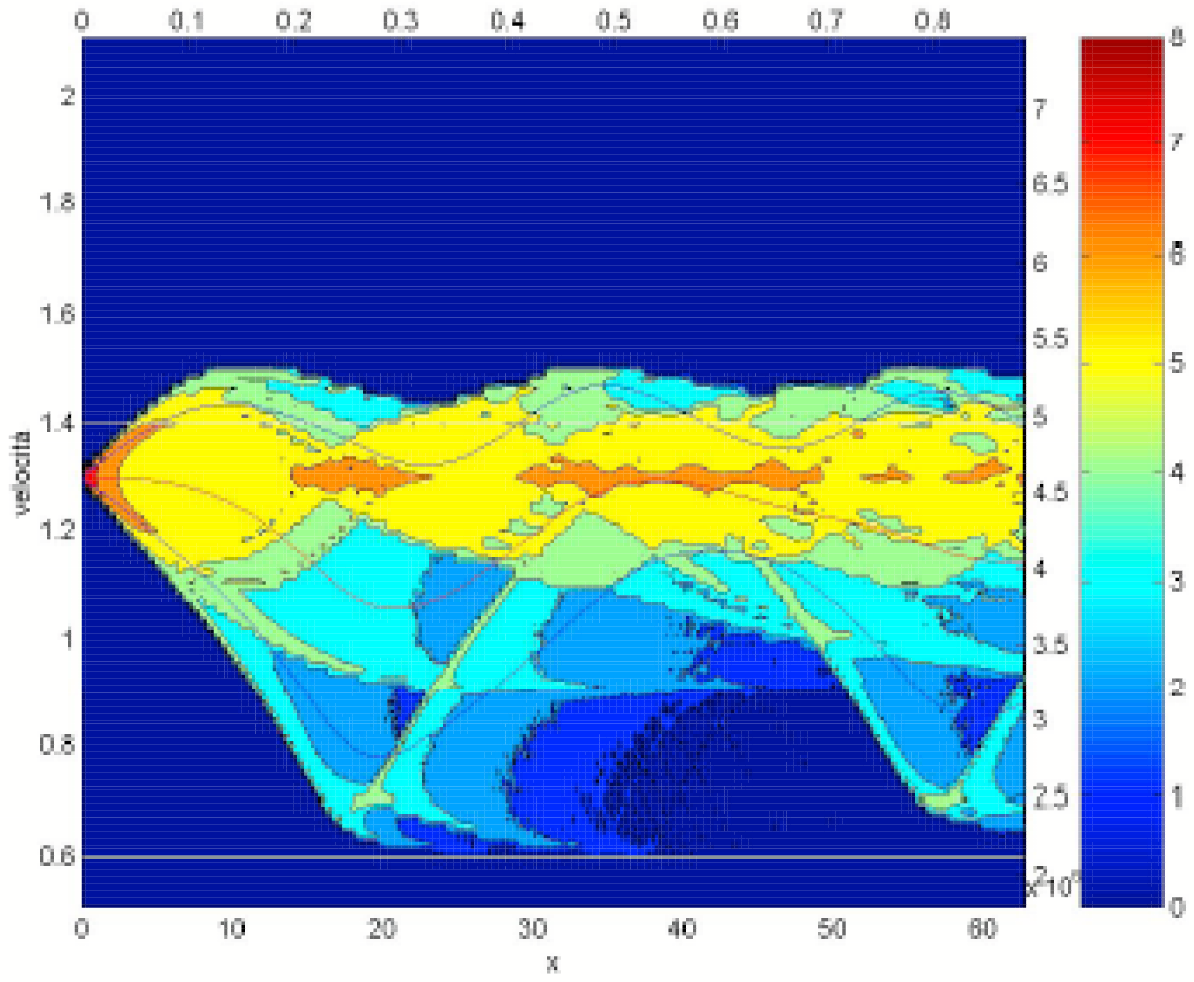}
 \end{center}
  \caption{Velocity distribution function of particles along the $x$ axis,
  when injected at $v_\rmb = 1.6$ (upper left), $v_\rmb = 1.5$ (upper right),
  $v_\rmb = 1.4$ (lower left) and $v_\rmb = 1.3$ (lower right),
  above the wave phase velocity, $v_\phi = 1$, with $m = k = 1$
  and wave amplitude $A = 0.04$. The cat's eye boundaries lie at
  $v_\phi + 2 \sqrt{A} = 1.4$ and $v_\phi - 2 \sqrt{A} = 0.6$.}
  \label{Lfig0411}  
\end{figure}

For particles injected with a velocity outside the wave cat's eye, the beam is
modulated. A significant difference between $(x,v)$ plots for propagating beams
and the
more familiar $(x,v)$ Poincar\'e sections of particle-in-wave dynamics
(see e.g.\ Ê\citep{EscDov81,Esc85,EEbook})
is the asymmetry between faster and slower particles,
obvious in Figure~\ref{Lfig0411}.

In particular, for particles injected with a velocity above the wave cat's eye,
the beam is moderately asymmetric. But
for a particle injection velocity within the capture range
  $[v_\phi - 2 \sqrt{A/m}, v_\phi + 2 \sqrt{A/m}]$,
the picture gets strongly deformed, as part of the beam is trapped as in
figure~\ref{Lfig0403}
while part of it moves outside the wave cat's eye.

\section{Particle dynamics in two waves : resonance overlap and chaos}
\label{secTwoWaves}

\begin{figure}
 \begin{center}
  \includegraphics[width=5cm,height=3.5cm]{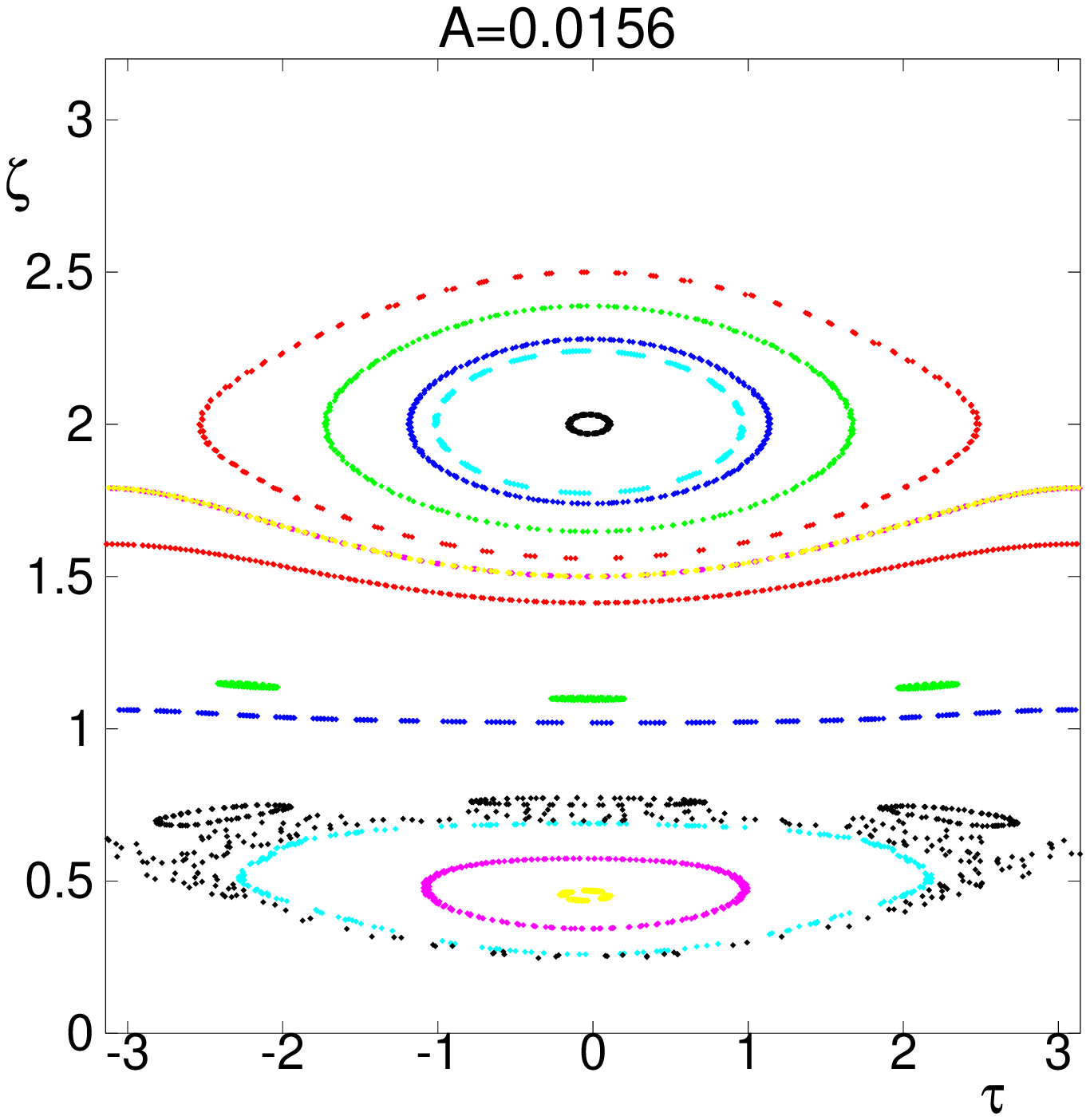}
  \includegraphics[width=5cm,height=3.5cm]{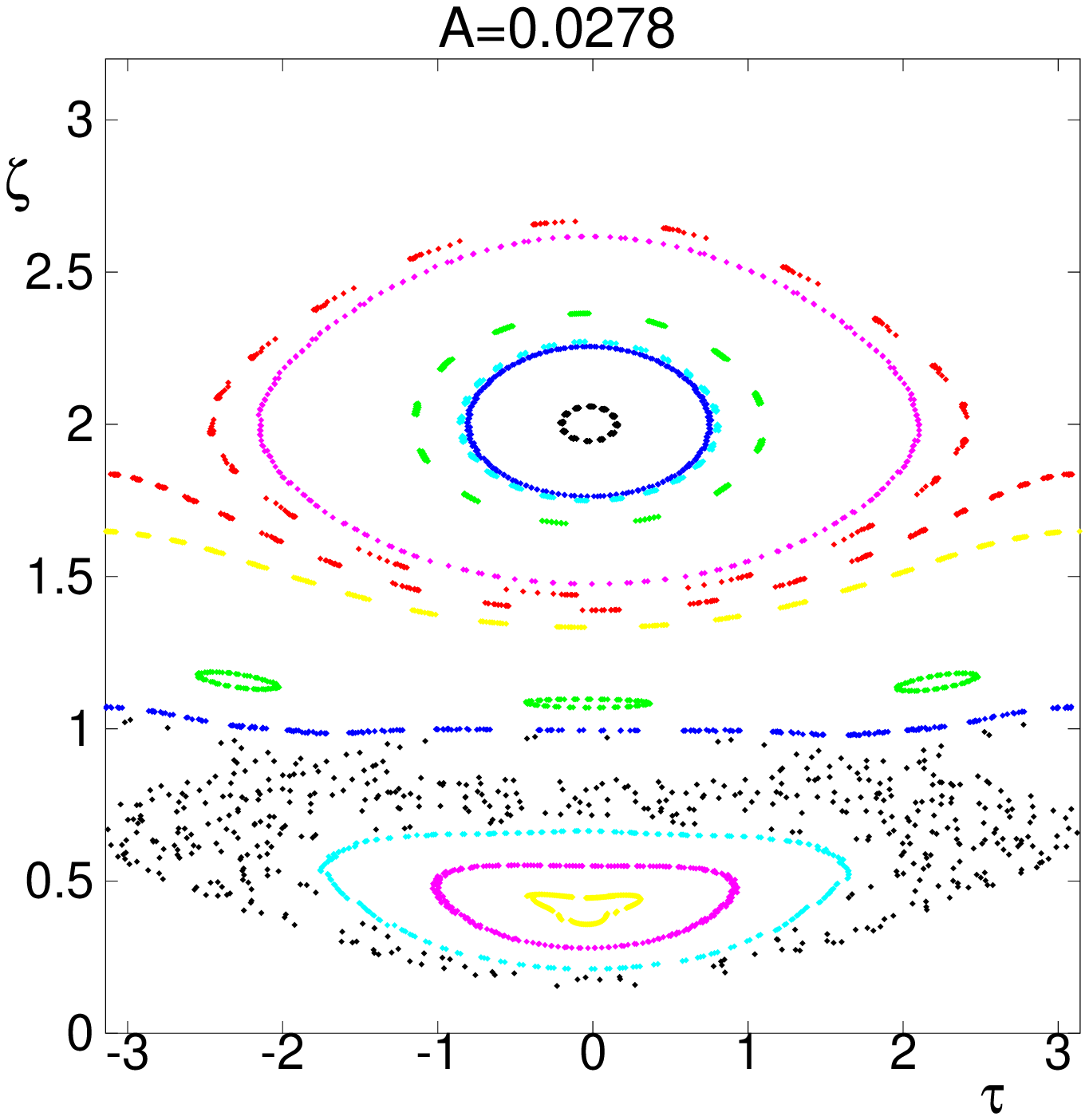}
  \includegraphics[width=5cm,height=3.5cm]{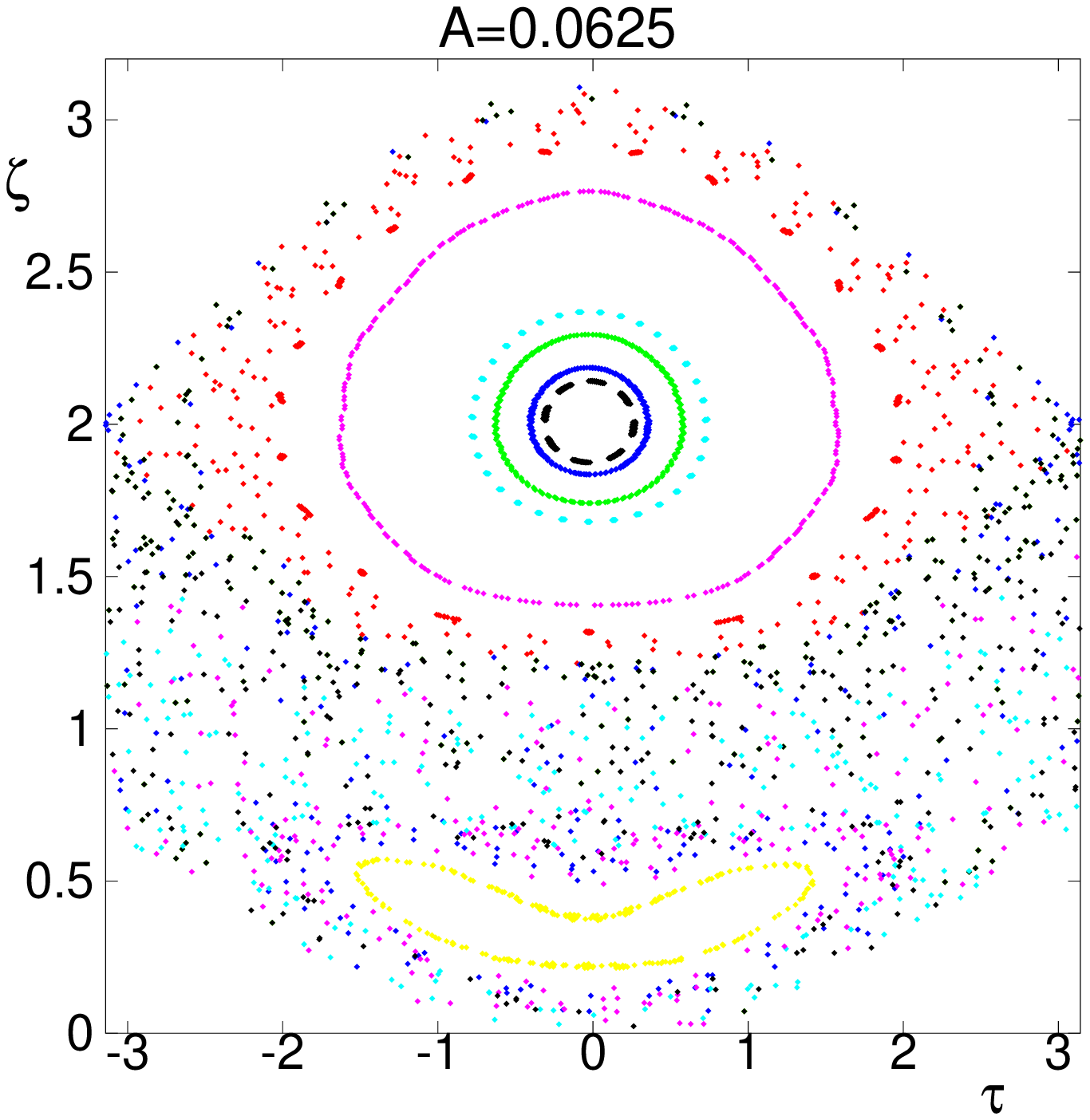}
 \end{center}
  \caption{Poincar\'e section in $(\tau, \zeta)$ variables, of the return
    map after one wavelength $L$, for 15 particles, with mass $m=1$.
    Wave parameters are $k_1 = 1$, $k_2 = 1/2$, $v_{\phi 1} = 1$, $v_{\phi 2} =
2$,
    $\phi_1 = \phi_2 = \pi$. Wave amplitudes $A_1 = A_2$ yield
    overlap parameter $s=0.5$ (left), 0.66 (centre) and 1 (right).}
  \label{Lfig0502}  
\end{figure}

The motion of a particle in the field of two waves is a paradigm of hamiltonian
chaos.
In our formulation, Poincar\'e sections are given by $x \, \mathrm{mod} \, L =
0$,
and the return map is symplectic, hence area-preserving in conjugate variables
$(\tau,
\zeta)$. Figure~\ref{Lfig0502} displays this Poincar\'e section, showing the
growth of
the chaotic domain for increasing wave amplitudes, and the destruction of KAM
tori
\citep{EscDov81}.

\begin{figure}
 \begin{center}
  \includegraphics[width=7cm,height=4.3cm]{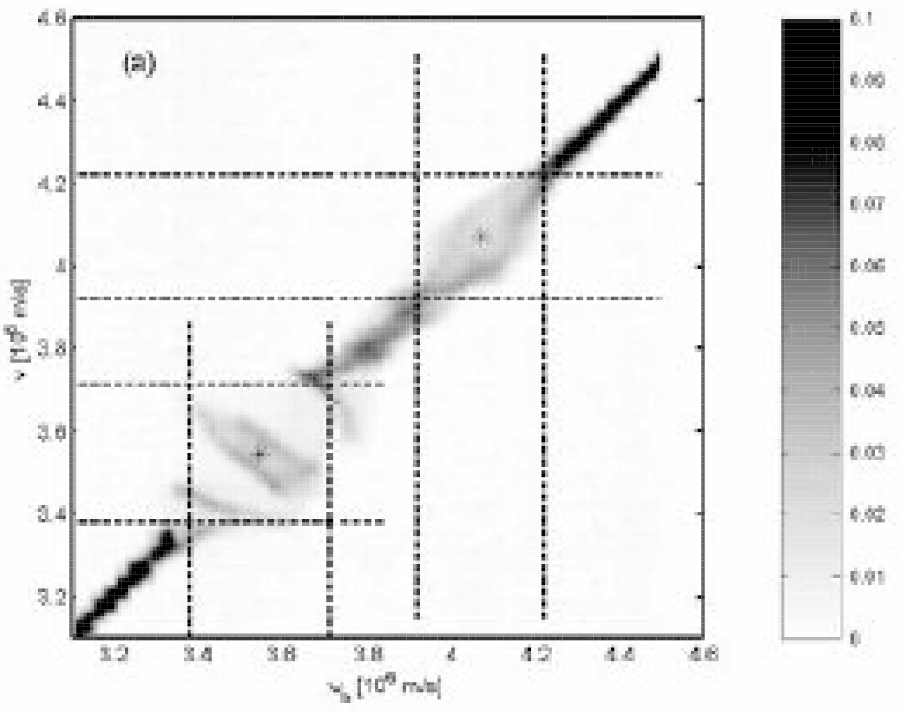}
  \includegraphics[width=7cm,height=4.3cm]{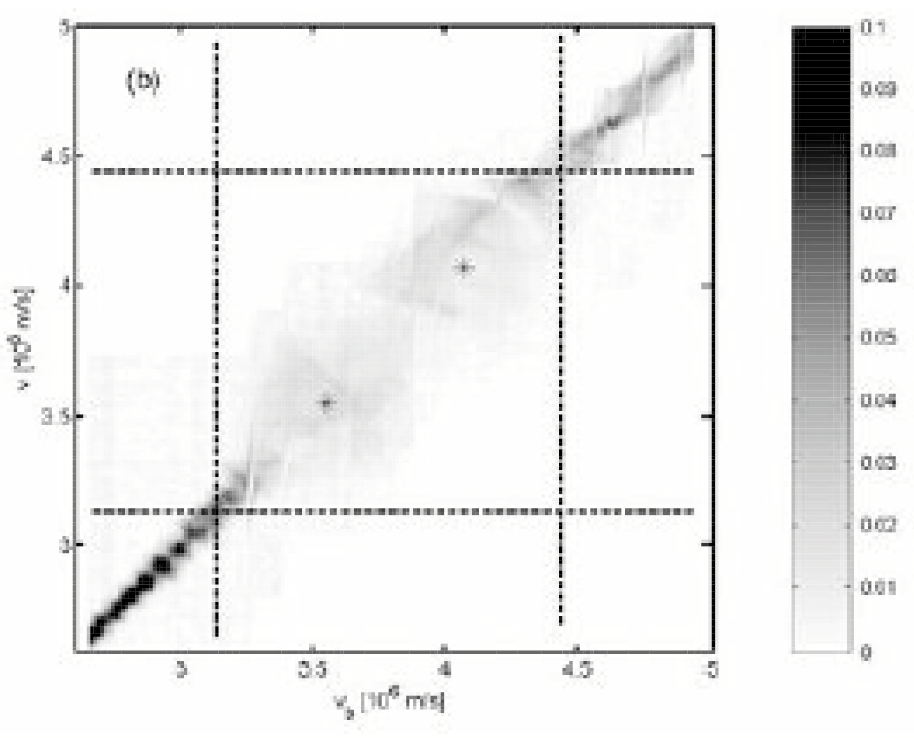}
  \includegraphics[width=7cm,height=4.3cm]{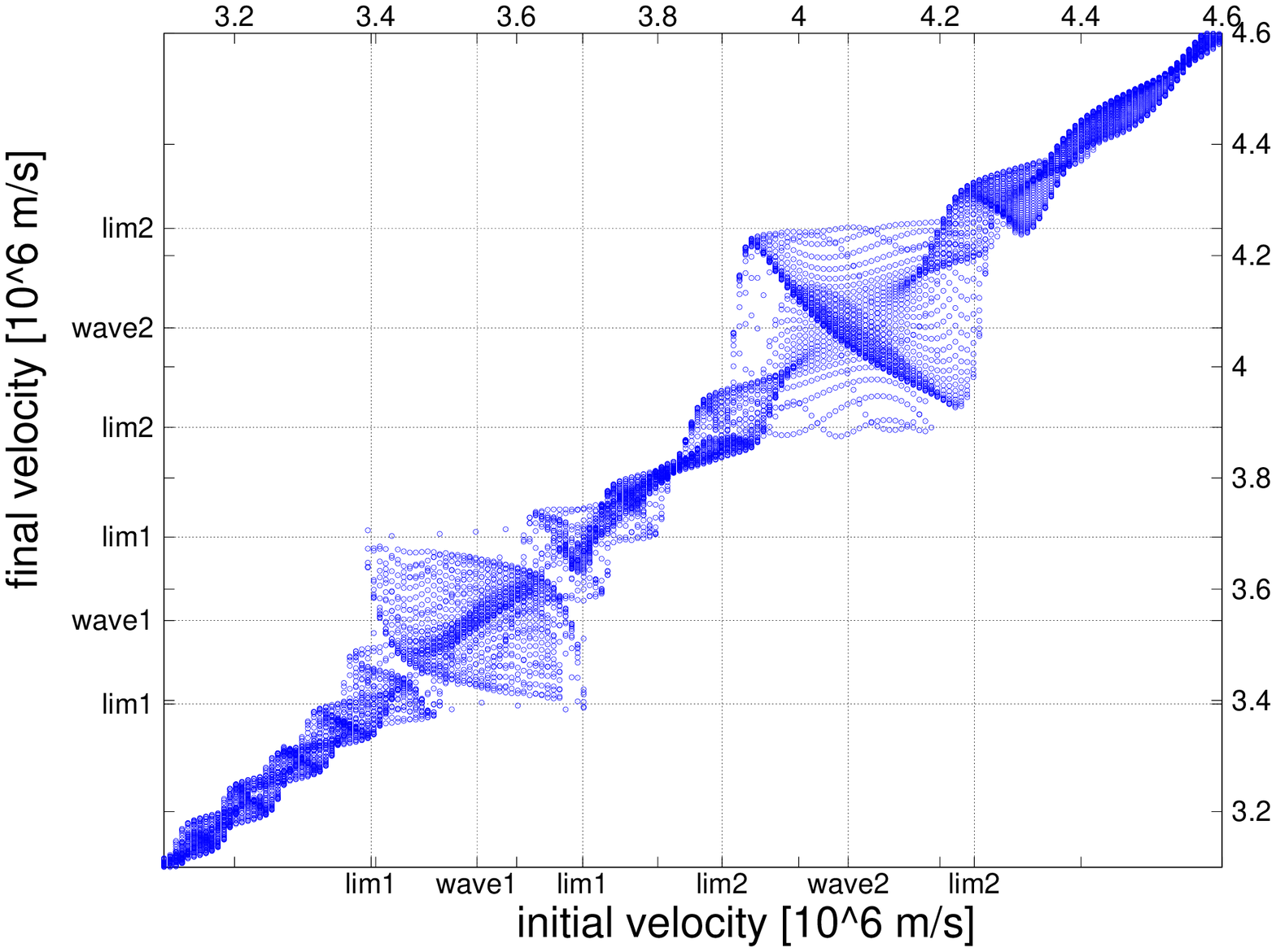}
  \includegraphics[width=7cm,height=4.3cm]{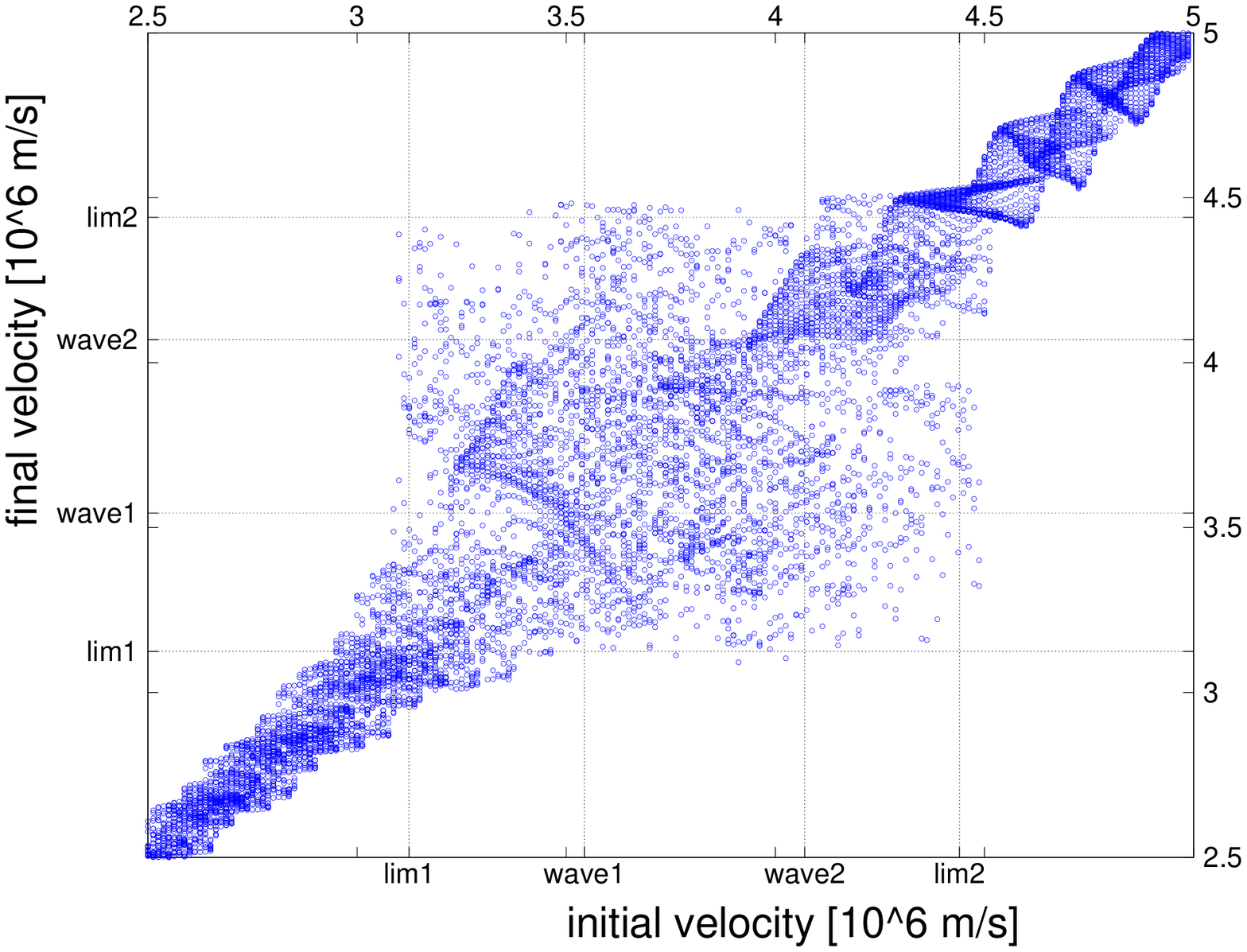}
\end{center}
  \caption{Velocities of particles after interaction with two waves, versus
injection velocity.
    (top) Experimental data with dotted lines marking cat eyes boundaries ;
    (bottom) numerical results.
    (left) Non-overlapping cat  eyes, $s = 0.63$ ;
    (right) overlapping trapping domains, $s = 1.5$.}
  \label{Lfig050304}  
\end{figure}

The corresponding transition to large scale chaos by increasing the resonance
overlap
parameter $s = 2 (\sqrt{A_1} + \sqrt{A_2}) / (|v_{\phi 2} - v_{\phi 1}|
\sqrt{m})$ is
also observed by recording the particle velocities at a fixed traveled distance
$L_0$,
after being injected at a fixed velocity $v_\mathrm{in}$. As seen in
figures~\ref{Lfig0411}, a cold beam injected in a wave cat's eye spreads over
the
velocity interval spanned by this cat eye, and if the beam is injected outside
cat eyes
it remains confined between the velocities of KAM tori on either side. Beam
velocity spreading (also called heating) has been used to diagnose resonance
overlap,
and our numerical scheme is compared with experimental data \citep{DovAuhMacGuy}
in
figure~\ref{Lfig050304}.

\begin{figure}
 \begin{center}
  \includegraphics[width=7cm,height=4.3cm]{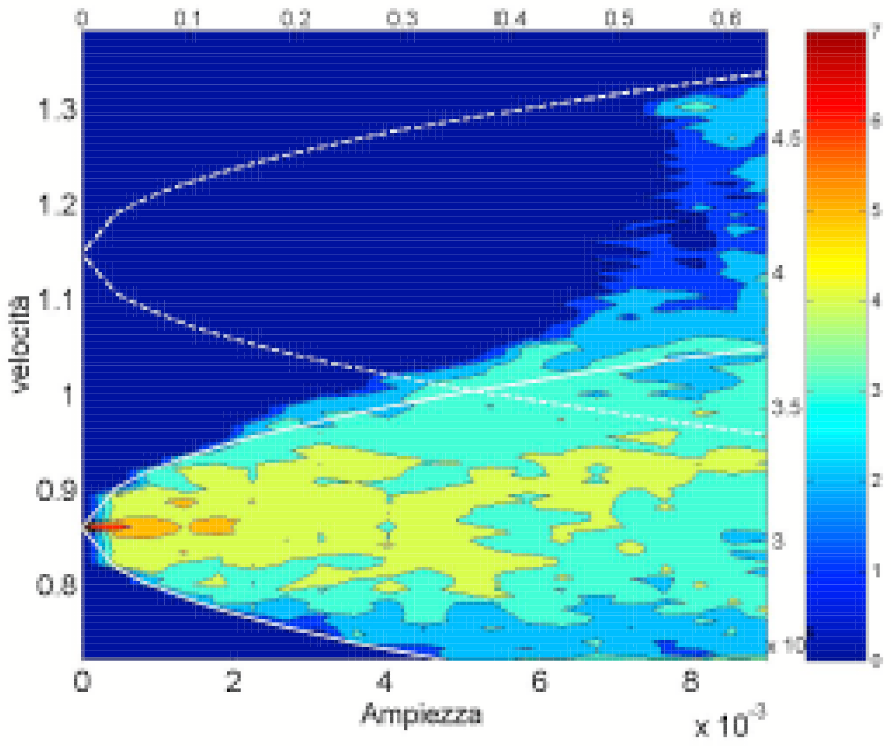}
  \includegraphics[width=7cm,height=4.3cm]{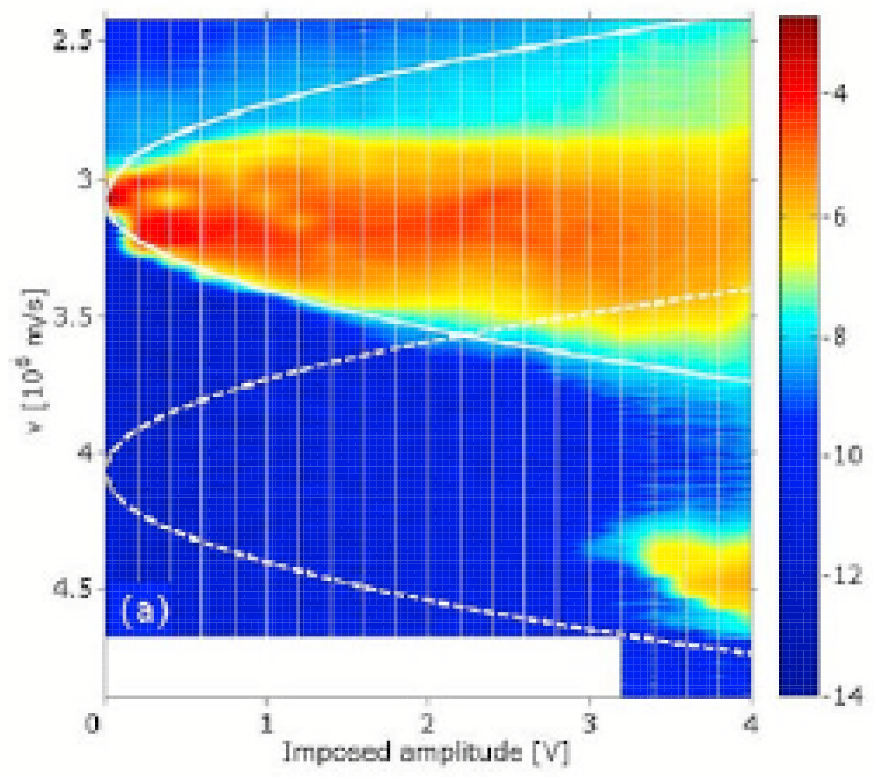}
\end{center}
  \caption{Velocities of particles after interaction with two waves,
    versus amplitude of waves $A_1 = A_2$.
    Cold beam injected at phase velocity of one wave.
    (left) Numerical results for $v_\rmb = v_{\phi 1} = 1.1487$,
           $v_{\phi 2} = 0.8609$, $k_1 = 0.6529$, $k_2 = 1.7424$.
    (right) Experimental data.}
  \label{Lfig0505}  
\end{figure}

Moreover, the transition to large scale chaos in phase space is known to occur
stepwise.
For increasing wave amplitudes, successive KAM tori get destroyed, so that the
beam
invades velocity domains resulting from the merging of capture regions
corresponding to
``secondary'' resonances \citep{EscDov81}. The accessible velocity interval for
the beam
injected in one wave then grows like a devil's staircase, the higher steps
corresponding
to the merging with major secondary resonances. Figure~\ref{Lfig0505} compares
these
domains obtained both numerically and experimentally \citep{MDE,DME,Buchanan}.
While
experimental data are blurred due to recording accuracy, numerical data have
limited
resolution due to the large number of particles (here only 25000) needed for the
sharp
observation of a threshold. Nevertheless, the agreement is quantitatively
satisfactory.

\section{Conclusion}
\label{secPersp}

The scheme has proved its relevance to describe accurately particle motion in a
given
field, along a single space dimension. It also provides new pictures of known
behaviours, which complement more familiar, usually more symmetric plots in
$(x,v)$
space.

Extending our approach to three space dimensions is rather straightforward provided the
particles stream along a given coordinate, say $x$. Higher-order symplectic schemes can
also be constructed by composing several maps $\cF_{\gamma_i \Delta x}$ and
$\cF_{\gamma_i \Delta x}^*$ with appropriate substeps $\gamma_i$ \cite{Hairer}. More
challenging is the issue raised by the particle feedback on the wave field, calling for
a self-consistent space-based model of particles and waves evolution, in the spirit of
models used for weak plasma turbulence \citep{EEbook}.

It will also be interesting to apply this scheme to model the many-waves regime
of weak
plasma turbulence, where particle velocity undergoes a chaotic transport over a
wide
range \citep{CEV,BE1,EEbook,Els10,BEEB}, as properties of this transport are
still
controversial.

\section{Acknowledgements}
\label{secThanks}

The work of A.~Ruzzon was made possible by an Erasmus mobility
from Universit\`a degli studi di Padova to universit\'e de Provence.
This work benefited from fruitful discussions with D.F.~Escande.




\appendix
\section{Explicit derivation of \refeq{eqsigma}}
\label{appA}

To reduce (\ref{can1cube}) to (\ref{eqsigma}), note that the derivatives of
momentum (\ref{cP2}) read
\begin{eqnarray}
  \partial_\zeta \cP (\tau, \zeta, x)
  & = &
  m \cP^{-1} \, ,
  \\
  \partial_\tau \cP (\tau, \zeta, x)
  & = &
  - m \cP^{-1} \partial_\tau V \, ,
\end{eqnarray}
and consider the dimensionless (both physically and numerically relevant)
quantity
$\sigma = (\tilde \zeta - \zeta) / (\zeta - V)$, which characterizes the changes
in particle energy per step and must be small for an accurate calculation.

Equation \refeq{map1zeta} reads, with arguments $(\tau, \tilde \zeta, x)$ in
$\cP$ and
$(\tau,x)$ in $V$,
\begin{eqnarray}
  \sigma
  & = &
  - \partial_\tau \cP \Delta x / (\zeta - V)
  \\
  & = &
  - \sqrt{\frac m 2} \, \partial_\tau V \Delta x \, (\tilde \zeta - V)^{-1/2}
(\zeta - V)^{-1}
  \\
  & = &
  - a \, \Bigl( \frac {\tilde \zeta - V}{\zeta - V} \Bigr)^{- 1/2}
  \\
  & = &
  - a \, \Bigl(1 + \frac {\tilde \zeta - \zeta}{\zeta - V} \Bigr)^{-1/2} \, ,
\end{eqnarray}
using the dimensionless $a$ defined by \refeq{perta}.
The fixed point equation \refeq{eqsigma} then follows.

\section{Explicit flowchart of the second order algorithm}
\label{appB}

The symmetric scheme \refeq{map2} may be expanded as follows:
\begin{itemize}
\item{$(\tau, \zeta, x) \mapsto (\tau, \zeta^*, x)$ :
  solve $\zeta
  =
  \zeta^*
  + \partial_\tau \cP (\tau, \zeta^*, x) \Delta x/2$
  for $\zeta^*$ ; }
\item{$(\tau, \zeta^*, x) \mapsto (\tau^*, \zeta^*, x)$ :
  $\tau^*
  =
  \tau
  + \partial_\zeta \cP (\tau, \zeta^*, x) \Delta x/2$ ; }
\item{$(\tau^*, \zeta^*, x) \mapsto (\tau^*, \zeta^*, \tilde x)$ :
  $\tilde x = x + \Delta x$ ; }
\item{$(\tau^*, \zeta^*, \tilde x) \mapsto (\tilde \tau, \zeta^*, \tilde x)$ :
  solve $\tau^*
  =
  \tilde \tau - \partial_\zeta \cP (\tilde \tau, \zeta^*, \tilde x) \Delta x/2$
  for $\tilde \tau$ ; }
\item{$(\tilde \tau, \zeta^*, \tilde x) \mapsto (\tilde \tau, \tilde \zeta,
\tilde x)$ :
  $\tilde \zeta
  =
  \zeta - \partial_\tau \cP (\tilde \tau, \zeta^*, \tilde x) \Delta x/2$.}
\end{itemize}


\begin{thebibliography}{99}

\bibitem
{BE1}
  D.~B\'enisti and D.F.~Escande,
  {Origin of diffusion in Hamiltonian dynamics},
  \textit{Phys. Plasmas} \textbf{4} (1997) 1576--1581.

\bibitem%
{BEEB}
  N.~Besse, Y.~Elskens, D.F.~Escande and P.~Bertrand,
  {Validity of quasilinear theory~: refutations and new numerical confirmation},
  \textit{Plasma Phys. Control. Fus.}Ê \textbf{53} (2011) 025012.

\bibitem
{Buchanan}
  M.~Buchanan,
  Richness in simplicity,
  \textit{Nature Physics} \textbf{2} (2006) 429.

\bibitem
{CEV}
  J.R.~Cary, D.F.~Escande and A.D.~Verga,
  {Non quasilinear diffusion far from chaotic threshold},
  \textit{Phys. Rev. Lett.} \textbf{65} (1990) 3132--3135.

\bibitem
{DovAuhMacGuy}
  F.~Doveil, Kh.~Auhmani, A.~Macor and D.~Guyomarc'h,
  Experimental observation of resonance overlap responsible for Hamiltonian
chaos,
  \textit{Phys. Plasmas} \textbf{12} (2005) 010702.

\bibitem
{DER}
  F.~Doveil, Y.~Elskens and A.~Ruzzon,
  Observation and control of hamiltonian chaos in wave-particle interaction,
  in : A.~Sen, S.~Sharma and P.N.~Guzdar (Eds),
  \textit{International symposium on waves, coherent structures and turbulence
in plasmas}
  (Gandhinagar, 2010),
  \textit{AIP Proc.} \textbf{1308} (2010) 132--141.

\bibitem
{DME}
  F.~Doveil, A.~Macor and Y.~Elskens,
  Direct observation of a ``devil's staircase'' in wave-particle interaction,
  \textit{Chaos} \textbf{16} (2006) 033103.

\bibitem
{Els10}
  Y.~Elskens,
  {Nonquasilinear evolution of particle velocity in incoherent waves
  with random amplitudes},
  \textit{Commun. Nonlinear Sci. Numer. Simul.} \textbf{15} (2010) 10--15.

\bibitem
{EEbook}
  Y.~Elskens and D.~Escande,
  \textit{Microscopic dynamics of plasmas and chaos}
  (Bristol: IoP Publishing, 2003).

\bibitem
{Esc85}
  D.F.~Escande,
  {Stochasticity in classical hamiltonian systems~:
  universal aspects},
  \textit{Phys. Rep.} \textbf{121} (1985) 165--261.

\bibitem
{EscDov81}
  D.F.~Escande and F.~Doveil,
  {Renormalization method for the onset of stochasticity in a
  hamiltonian system},
  \textit{Phys. Lett. A} \textbf{83} (1981) 307--310.

\bibitem
{Hairer}
  E.~Hairer, C.~Lubich and G.~Wanner,
  \textit{Geometric numerical integration~:
  Structure-preserving algorithms for ordinary differential equations}
  (New York: Springer, 2002).

\bibitem
{Henon}
  M.~H\'enon,
  On the numerical computation of Poincar\'e maps,
  \textit{Physica D} \textbf{5} (1982) 412-414.

\bibitem
{Kartikeyan}
  M.V.~Kartikeyan, E.~Borie and M.K.A.~Thumm,
  \textit{Gyrotrons -- High power microwave and millimeter wave technology}
  (Berlin: Springer, 2004).

\bibitem%
{Leimkuhler}
  B.~Leimkuhler and S.~Reich,
  \textit{Simulating hamiltonian dynamics}
  (Cambrige: University press, 2005).

\bibitem
{MDE}
  A.~Macor, F.~Doveil and Y.~Elskens,
  Electron climbing a ``devil's staircase'' in wave-particle interaction,
  \textit{Phys. Rev. Lett.} \textbf{95} (2005) 264102.

\bibitem
{MK78}
  H.E.~Mynick and A.N.~Kaufman,
  Soluble theory of nonlinear beam-plasma interaction,
  \textit{Phys. Fluids} \textbf{21} (1978) 653--663.

\bibitem
{RuzzonTesi}
  A.~Ruzzon,
  \textit{Studio numerico della propagazione spaziale d'un fascio di test
  in un tubo a onde progressive e confronto con l'esperimento}
  (tesi di laurea, Universit\`a degli studi di Padova, Padua, 2009).

\end{thebibliography}
\end{document}